\documentclass[journal]{IEEEtran}
\usepackage{amsfonts}
\usepackage[linesnumbered,ruled]{algorithm2e}
\usepackage{algpseudocode}
\usepackage{amsfonts}
\usepackage{bm}
\usepackage{amsmath}
\usepackage{exscale}
\usepackage{relsize}
\usepackage{cite}
\usepackage{amssymb}
\usepackage{stfloats}
\usepackage{extarrows}
\usepackage{enumerate}
\usepackage{color,xcolor}
\usepackage{graphicx}
\usepackage{multirow,tabularx}
\usepackage{subfigure}
\usepackage{hyperref}

\usepackage{times,amsmath,color,amssymb,graphicx,geometry, epsfig,psfrag,subfigure}

\makeatletter

\newcommand{\Rmnum}[1]{\expandafter\@slowromancap\romannumeral #1@}
\newcommand{\tabincell}[2]{\begin{tabular}{@{}#1@{}}#2\end{tabular}}
\makeatother

\ifCLASSINFOpdf
\else
\fi
\begin{document}
%

\title{A GCICA Grant-Free Random Access Scheme for
M2M Communications in Crowded Massive MIMO Systems}

\author{\IEEEauthorblockN{Huimei Han$^{1}$, Lushun Fang$^1$,  Weidang Lu$^1$,  Wenchao Zhai$^2$, Ying Li$^3$,  Jun Zhao$^4$}

\thanks{
Huimei Han, Lushun Fang, and Weidang Lu are with College of Information Engineering, Zhejiang University of Technology, Hangzhou, Zhejiang, 310032, P.R. China, Wenchao Zhai is with College of Information Engineering, China Jiliang University, Hangzhou, Zhejiang, 310018, P.R. China,  Ying Li is with the State Key Lab of Integrated Services Networks, Xidian University, Xi'an, 710071, P.R. China,  and Jun Zhao is  with School of Computer Science and Engineering, Nanyang Technological University, Singapore. (Email: \{hmhan1215,\,2111903007,\,luweid\}@zjut.edu.cn, zhaiwenchao@cjlu.edu.cn,  yli@mail.xidian.edu.cn,  junzhao@ntu.edu.sg)}


}

\maketitle

\begin{abstract}
A high success rate of  grant-free random access scheme is proposed to support massive access  for machine-to-machine communications in massive multiple-input multiple-output  systems. This scheme  allows  active user equipments (UEs) to transmit their modulated uplink messages along with super pilots consisting of multiple sub-pilots   to a base station (BS). Then, the BS performs channel state information (CSI) estimation and uplink message decoding by utilizing a proposed graph combined clustering independent component analysis (GCICA)  decoding algorithm, and then employs the estimated CSIs to  detect active UEs by utilizing the characteristic of asymptotic favorable propagation of massive MIMO channel. We call this proposed scheme as GCICA based  random access (GCICA-RA) scheme. We analyze  the successful access probability,   missed detection probability, and uplink throughput of the GCICA-RA scheme.  Numerical results show  that, the GCICA-RA scheme significantly improves the successful access probability and uplink throughput,  decreases missed detection probability, and provides low  CSI estimation error at the same time.
\end{abstract}

\begin{IEEEkeywords}
  Grant-free random access, independent component analysis (ICA), massive MIMO,  M2M communications.
\end{IEEEkeywords}

%
\IEEEpeerreviewmaketitle

\section{Introduction}

\IEEEPARstart{T}{HE} machine-to-machine (M2M) communications are centered on the intelligent interaction of  user equipments (UEs) without human intervention,  which is the enabler for the Internet of Things (IoT) to achieve the envision of the ``Internet of Everything"~\cite{iot1,iot2,things,mobile}. In recent years, M2M communications develop rapidly and  have been applied to many scenarios, such as smart medical, smart vehicle, smart logistics, etc. Cisco visual networking index and forecast  predicts that there will be around 28.5 billion connected UEs by 2022~\cite{crowdue}. The massive multiple-input multiple-output (MIMO) technology, which achieves significant improvements in  energy  and spectral efficiency and can serve massive UEs in the same time-frequency resource, is well suited for   M2M communications~\cite{effi1,nora,M2Mbook}.

For  M2M communications in massive MIMO systems, random access  procedure  is a critical  step to initiate a data transmission~\cite{randomaccess,randomMIMO}.
Since  payload data is usually in small size  and the number of UEs in a cell is envisioned in the order of hundreds or thousands in M2M communications~\cite{mMTC1}, random access procedure utilized in the long term evolution (LTE) network may induce excessive signaling overhead and cannot support massive access~\cite{randomaccess,propagation}.

Researchers are exploring new random access schemes for M2M communications in massive MIMO systems, and the grant-based random access schemes  have been proposed in recent years. E.~{Bj\"ornson}~\textit{et~al.}~proposed a strongest-user collision resolution (SUCRe) scheme, which allocates  pilots to active UEs with largest channel gain among the contenders~\cite{SUCR}. However, the number of successful accessing UEs  decreases with the increase of the number of contenders~\cite{SUCRGBPA}. To improve the pilot resource utilization of the SUCRe scheme,  SUCR combined idle pilots access (SUCR-IPA) scheme was proposed in~\cite{SUCRIPA}, where the weaker  UEs randomly select idle pilots  to increase the number of successful accessing UEs.
  A user identity-aided pilot access scheme was proposed for massive MIMO with interleave-division multiple-access systems, where the interleaver of each UE is available at the BS according to the one-to-one correspondence between UE's identity number (ID) and its interleaver~\cite{IDPA}.  However, since the grant-based random access schemes require two handshake processes  between the base station (BS) and UEs, considering the small packet transmission in M2M communication,   such kind of  random access scheme will introduce heavy signaling overhead and low data transmission efficiency.

  To address this problem, the grant-free random access schemes have attracted much attention in recent years, which allow  active UEs to transmit their pilots and uplink messages  to the BS directly and perform activity detection, channel state information (CSI) estimation, and uplink message decoding in one shot.
%
 J. Ahn~\textit{et~al.}~proposed a Bayesian based random access scheme to detect  the UE's activity  and estimate the  CSI jointly by utilizing the expectation propagation algorithm, considering the  BS with one antenna~\cite{ce5}.  L. Liu~\textit{et~al.}~proposed a  approximate message passing (AMP) based grant-free scheme  to achieve the joint  activity detection and CSI estimation for massive \mbox{MIMO} systems~\cite{ce8}.
However,  this AMP-based grant-free  random access  scheme requires long pilot sequence to achieve better performance, resulting in heavy pilot overhead. To support massive connectivity with low {access} delay and overhead, a novel grant-free random access scheme was proposed to detect active UEs  and uplink message without CSI estimation in one shot~\cite{EICA}. However, the CSI estimation is needed for some M2M traffics for downlink  beamforming in the case of time-division duplex (TDD) model, such as the telemedicine surgery requiring the downlink control signal transmission.
 The above-mentioned grant-free random access schemes consider the single pilot structure. Jiang~\textit{et~al.}~proposed to concatenate multiple orthogonal sub-pilots into a super pilot sequence,  and  the super pilot sequence is utilized for activity detection and CSI estimation~\cite{MultiplePreambles}. Simulation results show that, compared to the single pilot structure, this super pilot structure can improve the number of successful UEs by utilizing their proposed detection approach.  However, with the increase of the number of active UEs, the probability that more UEs have the same super pilot sequence increases, resulting in low row rank pilot selection matrix and thus degrading the number of successful UEs. This motivates interesting researches on how to support massive access by utilizing the super pilot structure.


To achieve this goal, we propose a graph combined clustering  independent component analysis based  random access (GCICA-RA) scheme. This scheme utilizes the super pilot structure and mainly consists of two steps. During the first step, all the active UEs  concatenate their randomly selected sub-pilots to  obtain their super pilot sequences, and send their super pilot sequences and modulated uplink messages to the BS. During the second step, the BS performs CSI estimation and uplink message decoding jointly by utilizing a proposed GCICA decoding algorithm,  employs the estimated CSIs to  detect active UEs by utilizing the characteristic of  asymptotic favorable propagation of massive MIMO channel, and sends random access response (RAR) to active UEs.
  We analyze  the successful access probability,   missed detection probability, and uplink throughput of the GCICA-RA scheme.  Numerical results show  that, compared to the random access scheme proposed in{~\cite{MultiplePreambles}}, the GCICA-RA scheme significantly improves the successful access probability and uplink throughput,  decreases missed detection probability, and provides low  CSI estimation error at the same time.

%
 The \textbf{main contributions} of this paper are summarized as follows.
\begin{itemize}






   \item We propose a GCICA algorithm to perform CSIs estimation and uplink message decoding jointly. Specifically, based on the received sub-pilot signal, the BS first utilizes a successive interference cancellation (SIC) algorithm to estimate  CSIs of active UEs, and then decodes the uplink message  by utilizing the estimated CSIs. Then, the BS utilizes  a proposed  CICA algorithm  to  decode  uplink message  and estimate CSIs of UEs whose CSIs cannot be estimated by the SIC algorithm.
      Simulation results show that, compared to the detection method proposed in~\cite{MultiplePreambles}, the number of successful UEs can be improved even multiple UEs having the same super pilot sequence.


            \item For the SIC algorithm, we propose to employ the estimated CSIs without other overhead to  find   edges connecting to a recovered variable node, by utilizing the asymptotic favorable propagation  of massive MIMO systems. This is different from the existing method where  indexes of selected pilots should be inserted into the uplink message to find edges connected to a recovered variable node~\cite{SUCRGBPA}, which degrades data transmission efficiency.
\item  We  analyze the upper bound for the  successful
access probability and uplink throughput of the proposed GCICA-RA scheme, and derive the lower bound for the  missed detection probability of the proposed GCICA-RA scheme.
\end{itemize}

The remainder of this paper is organized as follows. Section II introduces system model and the proposed GCICA-RA scheme. Section III describes SIC and CICA algorithms in the the proposed GCICA-RA scheme. We present the performance analysis in Section IV.  Simulation results and the conclusion are given in Section V and VI, respectively.

\textbf{Notations} utilized throughout this paper are described in  Table~\ref{tab:notation}.
\begin{table}[htbp]
\scriptsize
\centering
 \caption{notations.}\label{tab:notation}

   \begin{tabular}{|c|c|}
\hline
\linespread{2}
Notations & Description\\
\hline
Italic letters& Scalars\\
\hline
 Boldface
lower-case & Vectors\\
\hline
Boldface
upper-case letters& Matrices\\
\hline
$(\cdot)^T$  and $(\cdot)^H$ &~ \tabincell{c}{ The transpose and conjugate transpose \\
 of  a vector or a matrix} \\
 \hline
`*'& ~Complex conjugate of a vector or a matrix\\
\hline
$\boldsymbol{x}_{i}$& The $i^{th}$ element of a vector $\boldsymbol{x}$\\
\hline
$\mathbb{R}$& The set of all real numbers \\
\hline
$\mathcal{C}\mathcal{N}( \mu, \sigma^2 )$& ~ \tabincell{c}{A circularly-symmetric complex Gaussian \\ distribution  with mean $\mu$  and variance $\sigma^2$}\\
\hline
$|| \cdot||$& The Euclidean norm of a vector\\
\hline
$| \cdot |$& The cardinality of a set\\
\hline
\text{arg}($d$)& The  phase of complex $d$\\
\hline
$\lfloor \cdot \rceil$& The rounding operation\\
\hline
$[\bm{x}]_n$&  The $n^{th}$ element of vector $\bm{x}$\\
\hline
$\bm{X}(i)$ & ~The $i^{th}$ column of matrix $\bm{X}$\\
\hline
\end{tabular}

\end{table}


%

\section{SYSTEM MODEL  and The proposed GCICA-RA Scheme}

\subsection{SYSTEM MODEL}
In this paper, we consider massive MIMO  communication systems in  time-division duplex (TDD) mode. There is a BS with  $M$ antennas and $K$ single-antenna UEs in a cell, where the number of active UEs in a random access procedure is $N_a$.

\begin{figure}[htbp]
	\centering
	\includegraphics[scale =0.30] {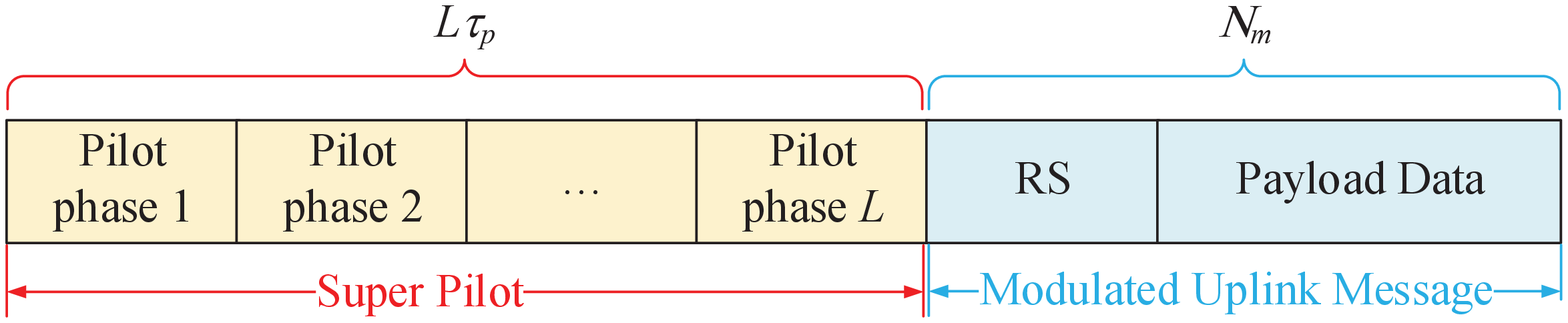}\\
	\caption{The frame structure of the proposed GCICA-RA scheme.}\label{frame-structure}
\end{figure}

 In the proposed GCICA-RA scheme,  each UE transmits its modulated uplink message and super pilot  to the BS directly. The BS performs CSI estimation and uplink message decoding by utilizing a proposed GCICA decoding algorithm, and then employs the estimated CSIs to  detect active UEs.
The frame structure of the GCICA-RA scheme is shown in Fig.~\ref{frame-structure}. Specifically, the super pilot of each UE consists of  $L$ consecutive sub-pilots, and the symbol length of each sub-pilot is $\tau_p$.
 Therefore, the symbol length of each super pilot is $L\tau_p$. In addition,   the proposed  GCICA decoding algorithm  utilizes multiple ICA classifiers to separate source signals. However, the phase of the separated signals is unpredictable.

  To solve this problem, each active UE  inserts a reference symbol (termed RS) with symbol length 1 into its uplink message~\cite{EICA}.  Let $\bm{v_k}=[v_k(1), v_k(2), \ldots, v_k({N_\text{m}})]^{{T}}$ denote the modulated uplink message of UE $k$, where $N_\text{m}$ represents the symbol length of the modulated uplink message. Then, $v_k(1)$, $[v_k(2),\ldots,v_k(N_\text{m})]$ represent the RS and the payload  data with symbol length $N_{\text{PD}}$,  respectively. Furthermore, since any complex symbol can be represented as a real symbol~\cite{sica}, we employ the BPSK modulation scheme in the proposed  GCICA-RA scheme,  i.e. $\upsilon_k(i)=\left\{ { \pm 1} \right\}$.

\begin{figure*}[htbp]
	\centering
	\includegraphics[scale =0.5] {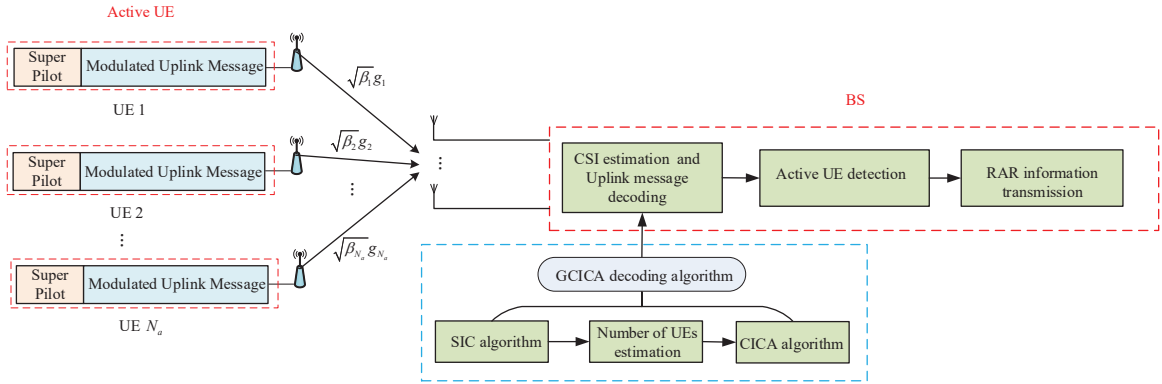}\\
	\caption{Block diagram of the proposed GCICA-RA scheme. Each active UE sends its super pilot and modulated uplink message to the BS. The BS performs CSI estimation and uplink message decoding by utilizing a proposed GCICA decoding algorithm,   employs the estimated CSIs to  detect active UEs, and sends RAR information to active UEs. In addition, the proposed GCICA decoding algorithm  utilizes a SIC algorithm to estimate CSIs  and thus decodes the uplink message of active UEs,  performs the number of active
UEs estimation, and utilizes a proposed CICA algorithm to decode uplink message and estimate CSIs
of  active UEs whose CSIs cannot be recovered by the SIC algorithm.
	}\label{system-model}
\end{figure*}

\subsection{ GCICA-RA scheme description}
As shown in  Fig.~\ref{system-model}, the proposed GCICA-RA scheme allows each UE to transmit its  modulated uplink message and super pilot to the BS directly. Then, the BS performs CSI estimation and uplink message decoding by utilizing a proposed GCICA decoding algorithm,   employs the estimated CSIs to  detect active UEs, and sends RAR information to active UEs. In addition, the proposed GCICA decoding algorithm first utilizes a SIC algorithm to estimate CSIs  and decode the uplink message of active UEs,  performs the number of active
UEs estimation, and utilizes a proposed CICA algorithm to decode uplink messages and estimate CSIs
of  active UEs whose CSIs cannot be recovered by the SIC algorithm.
The details are described as follows.


\subsubsection{ Step 1: Super pilot and modulated uplink message transmission}

\

There are $L$ sub-pilot phases, and  each active UE randomly  selects its sub-pilot from a set of mutually orthogonal normalized pilots $\bm{{{P}}_{{o}}}=\left\{ \bm{S_{1}},\bm{S_{2}}, \cdots,\bm{S_{\tau _p}} \right\} \in {\mathbb{R}^{{\tau _p} \times \tau_p}}$ during each sub-pilot phase. Concatenating these randomly selected sub-pilot sequences, each active UE obtains its super pilot. Following the frame structure as shown in Fig.~\ref{frame-structure}, each active UE sends it super pilot and modulated uplink message to the BS. In addition, the value of RS is the same for all active UEs, and we set the RS to be 1, i.e., $\upsilon_k(1)=1 \ \forall k$.



Then, let $\bm{g_k}={(g_k^1,g_k^2, \cdots ,g_k^M)^{{T}}}$ denote the \mbox{independent} and identically distributed (i.i.d.)  small-scale fading channel from UE $k$ to  the BS, i.e., $\bm{g_k}\sim \text{ }\mathcal{N}(0,{\bm{{{I}}_{M}}})$. Let  ${\rho_k}$  denote UE $k$'s transmitting power. Let  ${ {{\beta _k}} }$ stand for the channel gain between UE $k$ and the  BS, which can be known for UE $k$ before the random access procedure~\cite{SUCR}. Then, the received  sub-pilot signal during the $l^{th}\ (l = 1, 2, \ldots, L)$ sub-pilot phase at the  BS is
\begin{equation}\label{1}
\setcounter{equation}{1}
\bm{Y_{\text{p}}^{l}} = \sum\limits_{t = 1}^{{\tau_p}}\sum\limits_{k \in \bm{A_t^{l}}} {\sqrt {{\beta _k}{\rho_k}} }{{{\bm{g_k}(\bm{S_t})^{{T}}}}}  + \bm{Z_{\text{p}}^l},~l = 1, 2, \ldots, L,
\end{equation}
where $\bm{A_t^{l}}$ stands for the set of UEs that select pilot $S_t$ during the $l^{th}$ sub-pilot phase, $\bm{Z_{\text{p}}^l}\in {{\mathbb{R}}^{\text{  }M\times {\tau_p}}}$ denotes the additive white noise  matrix with each element following from a distribution of  $\mathcal{N}(0,{{\sigma }^{2}})$. In this paper, $\bm{Z}$ with different subscripts or superscripts will follow the same distribution. In addition,  we consider  ${{\rho_k}{{\beta }_{k}}} = 1$  to make the received signal from all  active UEs have the same power~\cite{POWER}. 

The received superimposed uplink message $\bm{Y_\text{m}} \in {{\mathbb{R}}^{\text{  }M\times {N_\text{m}}}}$ at the BS is
\begin{equation}\label{2}
\bm{Y_\text{m}} = \sum\limits_{k = 1}^{{N_a}} {{{\bm{g_k}(\bm{v_k})^{{T}}}}}  + \bm{Z_\text{m}}.
\end{equation}
\subsubsection{  CSI estimation, uplink message decoding, active UE detection and  RAR transmission} \label{SIC-d}

\

Based on the received sub-pilot signal  and the superimposed uplink message, the BS performs $ {\textcircled{1}}$ CSI estimation and uplink message decoding jointly, and  $ {\textcircled{2}}$ active UE detection and RAR information transmission. The details are described as follows.

\textbf{ $ {\textcircled{1}}$ CSI estimation and uplink message decoding}

The BS utilizes the  proposed GCICA  decoding algorithm to perform CSI estimation and uplink message decoding. The details of the GCICA  decoding algorithm are described as follows.

The BS first estimates the CSI corresponding to pilot $\bm{S_t}$ during the $l^{th}$ sub-pilot phase by utilizing  least squares (LS) method, denoted by $\bm{h_{t}^{l}}$
\begin{equation}\label{3}
\bm{h_{t}^{l}} = \bm{Y_{\text{p}}^{l}}\bm{(S_{t})^*} = \sum\limits_{k \in \bm{A_t^l}}{\bm{{g_k}}} + \bm{Z_{t}^{l}}.
\end{equation}
It can be seen from (\ref{3}) that,  during the $l^{th}$ sub-pilot phase,  the estimated CSI corresponding to  pilot  $\bm{S_t}$ is the sum of CSIs of active  UEs  selecting  pilot $\bm{S_t}$. So, considering all $L$ sub-pilot phases, a bipartite graph can be used to describe the proposed GCICA-RA scheme, and thus the SIC algorithm can be utilized to estimate UEs' CSIs on the bipartite graph. The procedure of the SIC algorithm is described in Section~\ref{SIC-procedure} in  detail. We use  $\bm{\hat{H}_{\text{SIC}}} =[\bm{\hat h_{\text{SIC}}^{1},\cdots,\hat h_{\text{SIC}}}^{N_{\text{SIC}}^{\text{s}}}]\in {\mathbb{R}}^{\text{  }M\times {N_{\text{SIC}}^{\text{s}}}}$ to denote the estimated CSIs  via the SIC algorithm, where $N_{\text{SIC}}^{\text{s}}$ is  the number of estimated CSIs  via the SIC algorithm. Thus, the detected uplink message can be obtained by utilizing LS method
\begin{equation}\label{4}
\bm{\hat{x}_{\text{SIC}}^i} = \frac{(\bm{\hat h_{\text{SIC}}^{i}})^{H} \bm{{Y_\text{m}}}}{(\bm{\hat h_{\text{SIC}}^{i}})^{H}(\bm{\hat h_{\text{SIC}}^{i}})},1 \leq i \leq N_{\text{SIC}}^{\text{s}}.
\end{equation}
After demodulating and decoding the  detected uplink message $\bm{\hat{x}_{\text{SIC}}^i}$, we obtain the corresponding decoded uplink message $\bm{\hat{v}_{\text{SIC}}^i}$.

Then, based on the estimated CSIs corresponding to  all pilots during the $l^{th}$ sub-pilot phase, the BS estimates the number of  active UEs by utilizing  the  estimation method  proposed in~\cite{EICA}, denoted by $\hat {{N_a}}$, which is given by
\begin{equation}\label{5}
\begin{array}{l}
\hat{{N_a}} =  {\sum\limits_{t=1}^{\tau_p}  \bm{\frac{(h_{t}^{l})^{H} h_{t}^{l}}{M}}} \to  {\sum\limits_{t=1}^{\tau_p} |\bm{A_t^l}|}
={N_a}, M \to \infty.
\end{array}
\end{equation}
We can observe from (\ref{5}) that  $\hat{{N_a}}$ approaches $N_a$ with large $M$. Therefore, the estimated number of active UEs whose CSIs cannot be obtained via the SIC algorithm is
\begin{equation}\label{6}
\hat{N}_\text{r} = \hat N_a - N_{\text{SIC}}^{\text{s}}.
\end{equation}

Finally, by utilizing $\hat{N}_\text{r}$, $\bm{\hat{v}_{\text{SIC}}^i} (1 \leq i \leq N_{\text{SIC}}^{\text{s}})$, and  $\bm{{Y_m}}$,   the BS employs a proposed CICA algorithm to obtain the CSIs and decoded uplink message of active UEs whose CSIs cannot be obtained via the SIC algorithm, denoted by  $\bm{\hat H_{\text{CICA}}} =[\bm{\hat h_{\text{CICA}}^{1}},\cdots,\bm{\hat h_{\text{CICA}}}^{{\hat{N}_{\text{r}}}}]\in {\mathbb{R}}^{M\times {\hat{N}_{\text{r}}}}$ and $\bm{\hat{v}_{\text{CICA}}^i}$, respectively. The details of the  CICA algorithm are descried in Section~\ref{CICA}.

\textbf{$ {\textcircled{2}}$ Active UE detection and RAR information transmission}


The BS detects an active UE by judging whether its  CSI can be estimated (i.e.,  whether the corresponding estimated CSI is a valid CSI estimation), and this UE realizes whether it has been detected by the BS in a distributed manner. The details are described as follows.

 We utilize $\hat{\bm{H}}=[\bm{\hat H_{\text{SIC}}}, \bm{\hat H_{\text{CICA}}}]$ to denote the estimated CSIs,
and assume that  $\hat{\bm{H}}(1)$ is a valid CSI estimation. More specifically,  $\hat{\bm{H}}(1)$ is the estimated CSI for  UE $m$ (i.e., $\hat{\bm{H}}(1)= \bm{g_m+\Delta g_m}$, where $\bm{\Delta g_m}$ denotes the estimation error and takes small  values). By correlating $\hat{\bm{H}}(1)$ with the  estimated CSI corresponding to  pilot $\bm{S_t}$  during the $l^{th}$ sub-pilot phase (i.e., $\bm{h_{t}^{l}},~t=1,2,\ldots,\tau_p$), we have
\begin{small}
\begin{equation}\label{7}
\begin{aligned}
\frac{(\bm{h_{t}^{l}})^{H} \hat{\bm{H}}(1)}{M}&=\frac{\left(\sum\limits_{k \in \bm{A_t^l}}{\bm{{g_k}}} + \bm{Z_{t}^{l}}\right)^{H}(\bm{g_m+\Delta g_m})}{M}\\
&=\frac{(\bm{{g_m}})^{H} \bm{g_m}}{M}+ \frac{\sum\limits_{k \in \bm{A_t^l},k\ne m}({\bm{{g_k}})^{H} \bm{g_m}}}{M} + \\
&  \frac{\sum\limits_{k \in \bm{A_t^l}}\bm{{(\bm{{g_k}})^{H}} \Delta g_m}}{M} + \frac{(\bm{Z_{t}^{l}})^{H} (\bm{g_m+\Delta g_m})}{M}\\
&\xlongrightarrow[\quad]{(a)}  1, M \to \infty,{\rm{ \text{if}}}~t~{\rm{ \text{is the index of}}}\\
&{\rm{ \text{the  pilot selected by UE}}}~m\\
\frac{(\bm{h_{t}^{l}})^{H} \hat{\bm{H}}(1)}{M}&=\frac{\sum\limits_{k \in \bm{A_t^l}}({\bm{{g_k}}})^{H}\hat{\bm{H}}(1)
+ (\bm{Z_{t}^{l}})^{H} \hat{\bm{H}}(1)}{M}\\
&\xlongrightarrow[\quad]{(b)} 0, M \to \infty, {\rm{otherwise}}
\end{aligned}
\end{equation}
\end{small}
where (a) follows from the fact that,  when $M$ goes into infinity, terms $\frac{(\bm{{g_m}})^{H}\bm{g_m}}{M}=1$, $\frac{\sum\nolimits_{k \in \bm{A_t^l},k\ne m}{\bm{{g_k}}^{H}}\bm{g_m}}{M}=0,  \frac{(\bm{Z_{t}^{l}})^{H}(\bm{g_m+\Delta g_m})}{M}=0$ by utilizing the asymptotic favorable propagation of massive MIMO channel,   term $\frac{\sum\nolimits_{k \in \bm{A_t^l}}{(\bm{{g_k}})^{H}} \bm{\Delta g_m}}{M}$ is close to $0$ since elements in $\bm{\Delta g_m}$ are small values, and (b) is  derived based on the random matrix theory. In addition, it is easy to derive that,  when  $\hat{\bm{H}}(i)$ is not a valid CSI estimation, $\frac{\left(\bm{h_{t}^{l}}\right)^{H}\hat{\bm{H}}(i)}{M}$ is close to 0 by utilizing the  random matrix theory.
Therefore,  the BS utilizes the following rule to judge whether $\hat{\bm{H}}(i)$ is a valid CSI estimation: for the $l^{th} \left(l \in \{ 1,...,L\}\right)$ sub-pilot phase, if there only exits an element in set $[\frac{(\bm{h_{l}^{1}})^{H}\hat{\bm{H}}(i)}{M},\cdots,\frac{(\bm{h_{l}^{t}})^{H}\hat{\bm{H}}(i)}{M},\cdots,\frac{(\bm{h_{l}^{\tau_p}})^{H}\hat{\bm{H}}(i)}{M}] $ close to 1,  $\hat{\bm{H}}(i)$ is a valid CSI estimation because each active UE only selects one pilot during each sub-pilot phase; otherwise, $\hat{\bm{H}}(i)$ is not a valid CSI estimation.


After obtaining all valid CSIs,  the BS generates and broadcasts the RAR information  $\bm{V}=\sum\nolimits_{u \in \bm{B}} (\bm{\hat{H}(u)})^{H}$ to all active UEs where $\bm{B}$ stands for the index set of valid CSI estimations.

The received RAR at UE $m$ is
\begin{equation}\label{8}
{R_m} = \sqrt {{\beta _m}} \bm{{V}{g_m}} + {Z_d}.
\end{equation}
Then, we have
\begin{equation}\label{9}
\begin{array}{l}
\frac{{{R_m}}}{{\sqrt {{\beta _m}} M}} = \frac{{\sqrt {{\beta _m}} \bm{{{V}}{g_m}}}}{{\sqrt {{\beta _m}} M}} + \frac{{{Z_d}}}{{\sqrt {{\beta _m}} M}} \vspace{1ex} \\
 = \frac{{\sqrt {{\beta_m}} \left( {\sum\limits_{u \in \bm{B}}  {{(\hat{\bm{H}}(u))}^H}{\bm{g_m}}} \right)}}{{\sqrt {{\beta _m}} M}} + \frac{{{Z_d}}}{{\sqrt {{\beta _m}} M}} \vspace{1ex} \\
\xlongrightarrow[\quad]{(a)M \to \infty}
\left\{ \begin{array}{l}
1{   \quad    {\text{\rm{if UE}}} ~m~ {\text{{\rm {has been detected}}}}}\\
\\
0{   \quad    \rm{otherwise}}
\end{array} \right.
\end{array}
\end{equation}
where (a) is derived by utilizing the asymptotic favorable propagation of massive MIMO channel, similar to (\ref{7}).
Based on the value of $\frac{{{R_m}}}{{\sqrt {{\beta _m}} M}}$, UE $m$ can realize whether it has been detected. If not,  UE $m$  will reattempt the random access procedure in the upcoming random access slot.

\section{ SIC and CICA algorithms}\label{algorithms}

\subsection{ SIC Algorithm}\label{SIC-procedure}

As we described in Section~II-B,  a bipartite graph
can be used to describe the proposed GCICA-RA scheme, and thus the SIC
algorithm can be utilized to estimate UEs' CSIs on the bipartite graph. In this subsection, we first present the bipartite graph representation, and then describe the SIC algorithm in detail.
\subsubsection{ Bipartite Graph Representation}

\

A bipartite graph  consists of variable nodes, factor nodes, and edges connecting  variable nodes and factor nodes. We utilize active UEs and available  pilots  during each sub-pilot phase to represent variable nodes and factor nodes, respectively. Furthermore,
if pilot $\bm{S_t}$ is selected by UE $k$ during sub-pilot phase $l$, we use $t^l$ to denote the corresponding factor node, and there exists an edge connecting the variable node $k$ with the factor node $t^l$. The CSI of UE $k$ is marked on  edges  connected to  UE $k$. Note that, since  we use the power control mechanism to make  ${{\rho_k}{{\beta }_{k}}} = 1$, we only need to estimate the small-scale fading channel. So, we only mark the small-scale fading coefficients between the BS and active UE. Furthermore, the degree of a node is referred to the number of edges connected to this node.

\begin{figure}[htbp]
	\centering
	\includegraphics[scale =0.45] {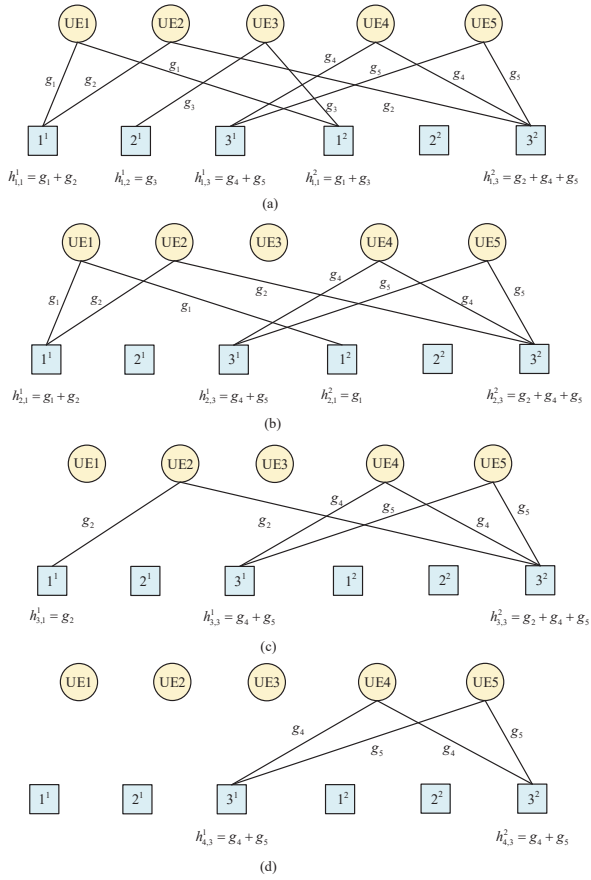}\\
	\caption{An example of SIC process. }\label{SIC process}
\end{figure}

 \textbf{Example:} Fig.~\ref{SIC process}(a) illustrates a bipartite graph based on 5 active UEs, 2 sub-pilot phases, and 3 available pilots. During each sub-pilot phase, active UEs (represented by  circles) are variable nodes and all available pilots (represented by  squares) are factor nodes. Since UE 1 and UE 2  select pilot $\bm{S_1}$ during sub-pilot phase $1$, UE 1 and UE 2 are connected to factor node $1^1$ by a black solid line. Since UE 3 selects pilot $\bm{S_2}$ during sub-pilot phase $1$, UE 3 is connected to factor node $2^1$ by a black solid line. Since UE 4 and UE 5  select pilot $\bm{S_3}$ during sub-pilot phase $1$, UE 4 and UE 5 are connected to factor node $3^1$ by a black solid line. Since UE 1 and UE 3 select pilot $\bm{S_1}$ during sub-pilot phase $2$, UE 1 and UE 3 are connected to factor node $1^2$ by a black solid line. Since UE 2, UE 4, and UE 5 all select pilot $\bm{S_3}$ during sub-pilot phase $2$, UE 2, UE 4 and UE 5 are connected to factor node $3^2$ by a black solid line.
\subsubsection{ SIC Algorithm}\label{SIC}

\

The BS employs SIC algorithm to estimate  UE's CSI on the bipartite graph, and then utilizes the estimated CSIs to decode the uplink message. In the following, 
we briefly discuss the SIC algorithm, present the pseudo-code of the SIC algorithm in Algorithm 1, and give an example to show the  SIC algorithm intuitively.

By performing the  autocorrelation operation on CSI corresponding to each  pilot during   each  sub-pilot phase, we have
\begin{equation}\label{10}
\begin{aligned}
\frac{(\bm{h_{t}^{l}})^{H} \bm{h_{t}^{l}}}{M} &=  \frac{\sum\limits_{k \in  \bm{A_t^l}}(\bm{{g_k}})^H{\bm{{g_k}}} + \left({\sum\limits_{k \in \bm{A_t^l}}{\bm{{g_k}}} + \bm{Z_{t}^{l}}}\right)^{H}\bm{Z_{t}^{l}}}{M}\\
&\xlongrightarrow[\quad]{(a)}  |\bm{A_t^l}|, M \to \infty,
\end{aligned}
\end{equation}
where (a) can be easily derived by utilizing the characteristic of   asymptotic favorable propagation of massive MIMO channel.

(\ref{10}) shows that, with large $M$, $\frac{(\bm{h_{t}^{l}})^{H}\bm{h_{t}^{l}}}{M}$ approaches the number of active UEs selecting  pilot $\bm{S_t}$ during the $l^{th}$ sub-pilot phase. In other words, $\frac{(\bm{h_{t}^{l}})^{H}\bm{h_{t}^{l}}}{M}$  represents the degree of factor node $t^{l}$.

 During the first SIC iteration,  the BS first finds factor nodes with  degree 1 by utilizing (\ref{10}). A factor node with degree 1  indicates that there is only one UE selecting the corresponding sub-pilot and the corresponding CSI can be recovered. Then, the BS finds other selected sub-pilots of UEs whose CSIs are recovered. More specifically, for each CSI corresponding to factor node with degree larger than 1,  after subtracting each recovered CSI,   the BS utilizes (\ref{10}) to compute the new degree of this  factor node. If the new degree  is less than the previous degree, this means that the sub-pilot corresponding to this factor node is selected by active UE corresponding to this recovered CSI. We subtract the recovered CSI from the CSI corresponding to this factor node and update the new degree as the previous degree of this factor node. This is equivalent to  delete the edges between this active UE and its selected sub-pilots on the bipartite graph. Thus, a new  bipartite graph can be obtained, and SIC iteration continues until there is no factor node with degree 1.

Since each active UE selects its sub-pilots randomly, there exits the case of UE $k$ having no pilot collision during multiple sub-pilot phases. This results in the case of  the recovered multiple CSIs corresponding to the same UE. Luckily, by utilizing  the asymptotic favorable propagation of massive MIMO channel, we can find these CSIs from the recovered  CSIs and select one of them as the recovered CSI of this UE. For example, if estimated $\bm{\hat h_1}$ and $\bm{\hat h_2}$ correspond to the same active UE, by utilizing  the asymptotic favorable propagation of massive MIMO channel, $\frac{(\bm{\hat h_1})^{H}\bm{\hat h_2}}{M}$ is close to 1 with large $M$. Otherwise, $\frac{(\bm{\hat h_1})^{H}\bm{\hat h_2}}{M}$ is close to 0 with large $M$.
Finally, the recovered CSIs via the SIC algorithm are  $\bm{\hat H_{\text{SIC}}} =[\bm{\hat h_{\text{SIC}}^{1},\cdots, \hat h_{\text{SIC}}}^{N_{\text{SIC}}^{\text{s}}}]\in {\mathbb{R}}^{\text{  }M\times {N_{\text{SIC}}^{\text{s}}}}$.

\begin{algorithm}[!t]
	\SetKwInOut{KIN}{Input}
	\SetKwInOut{KOUT}{Output}
	\SetKwInOut{Return}{Return}
	\caption{The SIC Algorithm}
	\KIN {$\bm{h_{t}^{l}},(t = 1, 2, \ldots, \tau_p; l = 1, 2, \ldots, L)$}
	\KOUT {$\bm{\hat{H}_{\text{SIC}}}$: a set of  recovered CSIs}
	$\bm{h}_{1,t}^{l}=\bm{h}_{t}^{l},flag=1, \bm{\hat{H}_{\text{SIC}}}= \phi,i=1$\;
	\While{$(flag\ne 0)$}
	{
		$flag=0$\;
		\For{  all $\bm{h}_{i,t}^{l}, l\in \{ 1,2,\ldots, L\}, t\in \{ 1,2,\ldots, \tau_p\}$ }
		{			
			Compute the degree  of the  corresponding  factor node  by utilizing (\ref{10})\;
			\If {the degree is 1}
			{
				Set $flag=1$\;
				$\bm{\hat{H}_{\text{SIC}}}=\bm{\hat{H}_{\text{SIC}}} \cup \bm{h}_{i,t}^{l}$\;
				\For{  all $\bm{h}_{i,m}^{n}, m\in \{ 1,2,\ldots, L\}, n\in \{ 1,2,\ldots, \tau_p\}$ }
				{
					\If{ $\frac{(\bm{h}_{i,m}^{n})^{H}\bm{h}_{i,t}^{l}}{M}$ is less than the previous degree }
					{
						update the channel response: $\bm{h_{i+1,m}^{n}} = \bm{h_{i,m}^{n}} -\bm{h}_{i,t}^{l}$\;
					}
					\Else
					{
						update the channel response: $\bm{h_{i+1,m}^{n}} = \bm{h_{i,m}^{n}}$\;
					}
				}
			}
		}
		$i=i+1$\;
	}\
	\Return{After deleting the recovered CSIs corresponding to the same UEs in set  $\bm{\bm{H}_{\text{SIC}}}$ and selecting one of
them as the recovered CSI, we have $\bm{\hat {H}_{\text{SIC}}}=[\bm{\hat h_{\text{SIC}}^1, \cdots, \hat h_{\text{SIC}}^{N_{\text{SIC}}^s}}]$}
\end{algorithm}

Algorithm 1 gives the detailed iteration process of SIC
algorithm, where $\bm{h_{i,j}^k}$  to represent the  CSI  corresponding to  factor node $j^k$ during the $i^{th}$ iteration. 

To better understand the SIC algorithm, an example is given to illustrate the SIC algorithm.  The details are described as follows:

 During the first iteration (Fig.~\ref{SIC process}(a)), the BS finds that the degree of factor node $2^1$ is 1 based on (\ref{10}). So the BS can directly recover the  CSI of UE 3, and incorporates this CSI (i.e., $\bm{h_{1,2}^{1}}$) into the set $\bm{\hat{H}_{\text{SIC}}}$.
 Then, after subtracting $\bm{h_{\text{SIC}}}^1$ from CSIs corresponding to factor node with degree larger than 1,  the BS finds that the new degree of factor node $1^2$ is less than the previous degree according to (\ref{10}), and thus realizes that the other index of sub-pilot selected by this UE is 1 in the second sub-pilot phase. Subtracting  the recovered CSI from the CSI corresponding to factor node $1^2$ (i.e, deleting the lines connected to variable node UE 3), a new bipartite graph can be obtained.
 Same as above operation, the BS can recover the CSI of UE 1 during the second iteration (i.e., $\bm{h_{2,1}^{2}}$ Fig.\ref{SIC process}(b)) and the CSI of UE 2 during the third iteration (i.e., $\bm{h_{3,1}^1}$ Fig.\ref{SIC process}(c)).
 In the end, since there is no factor node with degree 1 (Fig.\ref{SIC process}(d)),  SIC ends.  Finally, by utilizing the  asymptotic favorable
propagation of massive MIMO channel, we find that  the recovered CSIs (i.e., $\bm{h_{1,2}^{1}}$, $\bm{h_{2,1}^{2}}$, and  $\bm{h_{3,1}^1}$) belong to different UEs. Hence, we have  $\bm{\hat h_\text{SIC}^1}=\bm{h_{1,2}^{1}}$, $\bm{\hat h_\text{SIC}^2}=\bm{ h_{2,1}^{2}}$, and  $\bm{\hat h_\text{SIC}^3}=\bm{h_{3,1}^1}$.

\subsection{ CICA algorithm}\label{CICA}
We can see from Fig.~\ref{SIC process} that, for these 5 active UEs, we can only recover  CSIs of UE 1, UE 2, and UE 3 by utilizing the SIC algorithm, and UE 4 and UE 5 fail to access the network. We propose a CICA algorithm to further obtain the CSIs and decoded uplink messages of UE 4 and UE 5. We briefly describe the CICA algorithm in Algorithm 2.  In the following, we use the term ``the remaining active UEs" to represent active UEs whose CSIs cannot be recovered by the SIC algorithm for simplify representation. The proposed CICA algorithm first $ {\textcircled{1}}$  utilizes multiple ICA classifiers to separate the source signals and obtain the decoded signals, and then $ {\textcircled{2}}$ employs a proposed clustering algorithm to obtain estimated CSIs and decoded uplink messages of the remaining active UEs. The details are described as follows.
\begin{algorithm}[!t]
	\SetKwInOut{KIN}{Input}
	\SetKwInOut{KOUT}{Output}
	\SetKwInOut{Return}{Return}
	\caption{CICA decoding algorithm}
	\KIN {$\bm{Y_\text{m}^{'}}=[\bm{Y_\text{m}}(1), \bm{Y_\text{m}}(2),\cdots, \bm{Y_\text{m}}(M)]^{{T}}$, $\hat{N}_r$}
	\KOUT {$\bm{\hat{v}_\text{CICA}}$: a set of decoded uplink message, $\bm{\hat H_{\text{CICA}}}$: a set of estimated CSIs\;
}
	${r}=N_\text{I}, \bm{\hat{v}_\text{CICA}}= \phi, \bm{\hat H_{\text{CICA}}} = \phi$\;
	\While{$(r\ne 0)$}
	{
		$\bm{f_i} \leftarrow$   Select $\hat{N}_r$ row vectors from $\bm{Y_{m}^{'}}$ randomly to obtain $\bm{ Y_\text{ICA}^{i}}$,  and then feed it to the $i^{th}$ ICA classifier to obtain  separated source signal $\bm{f_i} =[\bm{f_i^1},\bm{f_i^2},...,\bm{f_i^{\hat{N}_r}}]$\;
		$r= r-1$\;
	}
	\For{  all $i\in \{ 1,2,...,N_\text{I}\}, t\in \{ 1,2,...,{\hat{N}_r}\}$ }
	{
		$\bm{\hat{f_i^t}} \leftarrow$  Use the first decoding symbol in $\bm{f_i^t}$ to solve the ICA phase ambiguity problem via (\ref{12})\;
		$\text{DEC}(\bm{\hat{f_i^t}}) \leftarrow$ After the first symbol in $\bm{\hat{f_i^t}}$ is deleted, $\bm{\hat{f_i^t}}$ is fed into the decoder to obtain the decoded signal  $\text{DEC}(\bm{\hat{f_i^t}})$\;
		$j\leftarrow$ Judge which class $\text{DEC}(\bm{\hat{f_i^t}})$ belongs to by using the proposed clustering algorithm (assume it belongs to class $j$) \;
      $\bm{C_j}\leftarrow$ Incorporate this signal to the decoded signal set of the corresponding class \;
		$\bm{\hat d_j}\leftarrow$  Add the  remodulated signal  via (\ref{13});
	}
	\For{$j \in  \{ 1,2,...,{\hat{N}_r}\}$ }
	{
		$\bm{\hat{v}_\text{CICA}^j} = \text{sign}(\bm{\hat d_j})$, $\bm{\hat{v}_\text{CICA}}=\bm{\hat{v}_\text{CICA}}\cup \bm{\hat{v}_\text{CICA}^j}$\;
$\bm{\hat{h}_{\text{CICA}}^j} = \frac{\bm{{Y_m^{'}}({{\hat{\bm{v}}_\text{CICA}^j}})^{*}}}{({{\hat{\bm{v}}_\text{CICA}^j}})^H ({{\hat{\bm{v}}_\text{CICA}^j}})}$ \;
$\bm{\hat H_{\text{CICA}}} =\bm{\hat H_{\text{CICA}}} \cup \bm{\hat{h}_\text{CICA}^j}$ \;}
	\Return{$\bm{\hat{v}_\text{CICA} =[\hat{v}_\text{CICA}^1,\hat{v}_\text{CICA}^2,\cdots, \hat{v}_\text{CICA}^{\hat{N}_r}]}$\;
\ \ \ \ \ \ \ \ \ \ \ $\bm{\hat H_{\text{CICA}}} =[\bm{ \hat h_{\text{CICA}}^{1}},\cdots,\bm{ \hat h _{\text{CICA}}}^{{\hat{N}_{\text{r}}}}]$.}
\end{algorithm}

$ {\textcircled{1}}$ \textbf{ICA classifiers}

The BS first obtains the sum  of received superimposed uplink message of the remaining active UEs, denoted by $\bm{Y_\text{m}^{'}} \in \mathbb{R}^{M \times N_\text{m}}$.
\begin{equation}\label{11}
\begin{array}{l}
\bm{Y_\text{m}^{'}} = \bm{Y_\text{m}} - \sum\limits_{i=1}^{N_{\text{SIC}}^{\text{s}}}{\bm{h_{\text{SIC}}^{i}}(\bm{\hat{v}_{\text{SIC}}^i})^{T}}\\
          ={\sum\limits_{i = 1}^{{N_\text{r}}} {{{\bm{g_i}(\bm{v_i})^{\text{T}}}}}}  + \bm{Z_\text{m}^{'}}\\
           =\bm{G}_\text{r}\bm{V}_\text{r}+\bm{Z_\text{m}^{'}}.
\end{array}
\end{equation}
where ${N_\text{r}}$ denotes the number of the remaining active UEs. Recall that we
 can obtain the estimated value of ${N_\text{r}}$ (i,e., $\hat{N}_{\text{r}}$) by utilizing  (\ref{6}).

 Then, by utilizing the ICA method in~\cite{EICA}, the BS utilizes  $N_\text{I}$ ICA classifiers to separate  the source signals $\bm{V_\text{r}}$ in $\bm{Y_\text{m}^{'}}$. More specifically, the BS randomly selects $\hat{N}_{\text{r}}$  different row vectors from $\bm{Y_\text{m}^{'}}=[\bm{Y_\text{m}^{'}}(1),\cdots,\bm{Y_\text{m}^{'}}(M)]^{T}$ as the input signal of the $i^{th}$ ICA classifier, and the $i^{th}$ ICA classifier outputs the separated source signal, denoted by $\bm{f_i}=[\bm{f_i^{1}},\cdots,\bm{f_i^{\hat{N}_{\text{r}}}}],~i=1,\ldots, N_\text{I}$. We further use the first symbol in each separated source signal to solve the phase ambiguity problem introduced by the ICA classifier as follows
\begin{equation}\label{12}
\bm{\hat{f_i^t}}={f_i^1}(t)\times\bm{f_i^t}, (i\in \{ 1,2,...,N_\text{I}\}, t\in \{ 1,2,...,{\hat{N}_{\text{r}}}\}).
\end{equation}

Finally, after deleting the first symbol in  $\bm{\hat{f_i^t}}$, $\bm{\hat{f_i^t}}$ is fed into  the channel decoder  to obtain the decoded signal $\text{DEC}(\bm{\hat{f_i^t}})$. Thus, we can obtain the decoded signal from the $i^{th}$ ICA classifier, i.e., $\text{DEC}(\bm{f_i})=[\text{DEC}(\bm{f_i^{1}}),\cdots,\text{DEC}(\bm{f_i^{t}}),\cdots,\text{DEC}(\bm{f_i^{\hat{N}_{\text{r}}}})]$.

$ {\textcircled{2}}$ \textbf{Clustering algorithm}

The order of the ICA classifier output is undetermined, and thus we cannot sure which  UE  each  decoded signal from the $i^{th}$ ICA classifier belongs to. To solve this problem, we propose a clustering algorithm to cluster the decoded signals  of the same UE from these $N_\text{I}$ ICA classifiers, and further obtain the estimated CSIs and decoded uplink messages  of the remaining active UEs.

 We regard each remaining active UE as a class, and the decoded signals  of one UE belong to the same class. Therefore, there are ${\hat{N}_{\text{r}}}$ classes, indexed by $1,2,\ldots,{\hat{N}_{\text{r}}}$. Actually, since each ICA classifier outputs  decoded signals of ${\hat{N}_{\text{r}}}$ UEs, we can utilize  decoded signals from any ICA classifiers as the  first sample in each class. Here, we assume that ${\hat{N}_{\text{r}}}$ decoded signals from the first ICA classifier (i.e., $\text{DEC}(\bm{f_1})$) are the first sample in each class accordingly.


 The BS first judges which class each decoded signal from the $i^{th}~(i=2,\ldots,{\hat{N}_{\text{I}}})$ ICA classifier belongs to. More specifically,
considering that the fewer the number of different bits between two decoded signals, the more likely these two decoded signals come from the same UE.
The BS  compares the number of different bits between the decoded signal $\text{DEC}(\bm{f_i^{t}}),~(t=1,2,\ldots,{\hat{N}_{\text{r}}})$ and each element in $\text{DEC}(\bm{f_1})$, and judges  $\text{DEC}(\bm{f_i^{t}})$ belonging to the class which has the minimum number of different bits with  $\text{DEC}(\bm{f_i^{t}})$. We use $\bm{C_j}$ to denote a set of decoded signals in the $j^{th}$ class.

Then, for the $j^{th}$ class, the BS remodulates  each  element in set $\bm{C_j}$ by utilizing the  BPSK modulator and performs the sum operation as follows
\begin{equation}\label{13}
\bm{\hat d_j} = \sum\limits_{i=1}^{|\bm{C_j}|} 2\times {\bm{C_j}(i)}-1,
\end{equation}
where $2\times {\bm{C_j}(i)}-1$ refers  to the remodulated signal.

Finally, the decoded   uplink message of the remaining active UE can be  obtained by
\begin{equation}\label{14}
{{\hat{\bm{v}}_\text{CICA}^j}} = \text{sign}(\bm{\hat d_j}),~j=1,\ldots,{\hat{N}_{\text{r}}},
\end{equation}
and  the  estimated CSIs of the remaining active UEs can be obtained by
\begin{equation}\label{15}
\bm{\hat{h}_{\text{CICA}}^j} = \frac{\bm{{Y_m^{'}}({{\hat{\bm{v}}_\text{CICA}^j}})^{*}}}{({{\hat{\bm{v}}_\text{CICA}^j}})^H ({{\hat{\bm{v}}_\text{CICA}^j}})},~j=1,\ldots,{\hat{N}_{\text{r}}}.
\end{equation}

\section{ Performance Analysis}
In this section, we analyze the performance of the proposed GCICA-RA scheme, including the  successful access probability, missed detection  probability, and the uplink throughput.


\subsection{Successful access probability analysis}
Based on the procedure of the proposed GCICA-RA scheme, if the CSI of one UE can be obtained, this UE is a detected UE and its uplink message can also be decoded. This means that this UE can access the network successfully.
The GCICA-RA scheme first utilizes the SIC algorithm to obtain the CSIs of active UEs, and then employs the proposed CICA algorithm to obtain the CSIs of active UEs whose CSI cannot be recovered by the SIC algorithm.
 Let $S_\text{GCICA-RA}$, $S_\text{SIC}$ and $S_\text{CICA}$ denote the number of successful UEs obtained by  the proposed GCICA-RA scheme, the SIC algorithm and the CICA algorithm, respectively. Then, the  successful  access probability of the proposed GCICA-RA scheme is given by
\begin{equation}\label{16}
P_{s} =\frac{S_\text{GCICA-RA}}{N_a}=\frac{S_\text{SIC} +S_\text{CICA}}{N_a}.
\end{equation}
Next, we discuss how to compute terms $S_\text{SIC}$ and $S_\text{CICA}$ in (\ref{16}).

\subsubsection{$S_{\rm{SIC}}$ computation }\label{SIC number}

\

To compute $S_\text{SIC}$, we follow  the two assumptions below~\cite{SUCRGBPA}:

{\emph{Assumption 1}}: If an active UE has no pilot collision with other  active UEs, its CSI can be recovered successfully.


{\emph{Assumption 2}}: When an active UE selects the same pilot with other $l$ active UEs, if  CSIs  of these $l$ UEs can be recovered, its CSI can be recovered successfully.


Let $P_\text{fail}$ denote the  probability that UE's CSI cannot be recovered by the SIC algorithm. Then, we have
\begin{equation}\label{17}
S_\text{SIC}=N_a \left(1-P_\text{fail}\right).
\end{equation}
The and-or tree principle is used to study random process, and we  employ this principle to compute  $P_\text{fail}$. In  the following, we first derive the degree distributions of the proposed GCICA-RA scheme, and then discuss how to  utilize the  and-or tree principle to compute $P_\text{fail}$ based on the derived degree distributions.

\textbf{$ {\textcircled{1}}$ Degree distributions of the proposed GCICA-RA scheme derivation}

Let $\Lambda_l$ and $\Psi_l$ denote the probability that the degree of a variable node is $l$ and the probability that the degree of a factor node is $l$, respectively.  Then, the corresponding polynomial representations are given by
\begin{equation}\label{18}
\Lambda (x) \triangleq \sum\nolimits_l {\Lambda_l x^l},\quad \Psi (x) \triangleq \sum\nolimits_l {\Psi_l x^l}.
\end{equation}

The proposed GCICA-RA scheme allows each active UE  to select its sub-pilot during each sub-pilot phase, and there are $L$ sub-pilot phases. Therefore, the degree of each variable node is $L$, i.e.,
\begin{equation}\label{19}
\Lambda_{l} =\left\{
\begin{aligned}
1, & &l=L,  \\
0, & &{\rm{otherwise}}. \\
\end{aligned}
\right.
\end{equation}
In addition, since each active UE randomly selects its sub-pilot during each sub-pilot phase, the event that $l$ active UEs select the same sub-pilot
follows a Bernoulli distribution with parameters $N_a$ and ${1}/{\tau_p}$. Thus, the probability of a factor  node with degree $l$ (i.e., $\Psi_l$) is
\begin{equation}\label{20}
\Psi_l = \binom{N_a}{l} \left( \dfrac{1}{\tau_p}\right)^l \left(1- \dfrac{1}{\tau_p}\right)^{\left({N_a} - {l}\right)}.
\end{equation}

Let $\lambda_l$ denote the probability of an edge connecting  to a variable node with degree $l$, and $\rho_l$  denote the probability of an
edge connecting to a factor node with degree $l$. Then, we transfer the node-perspective presentations in (\ref{19}) and (\ref{20}) into the edge perspective presentations, respectively
\begin{equation}\label{21}
\lambda_l = \dfrac{\Lambda_l l}{\sum\nolimits_l {\Lambda_l l}}, \quad \rho_l = \dfrac{\Psi_l l}{\sum\nolimits_l {\Psi_l l}},
\end{equation}
and the polynomial representations are
\begin{equation}\label{22}
\lambda (x) \triangleq \sum\nolimits_l {\lambda_l x^{l-1}}, \rho (x) \triangleq \sum\nolimits {\rho_l x^{l-1}}.
\end{equation}

\textbf{$ {\textcircled{2}}$ $P_\text{fail}$ computation }
\begin{figure}[htbp]
	\centering
	\includegraphics[scale =0.3] {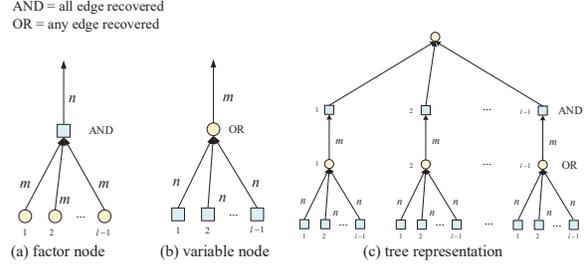}\\
	\caption{Probability updates}\label{tree}
\end{figure}

 Considering a factor node with degree $l$ (i.e., there are $l$ UEs selecting the corresponding pilot),  an edge connecting to  this factor node can be recovered when the other $l-1$ edges connecting to this factor node  have been recovered, as shown in Fig.~\ref{tree}(a). Similarly, as shown in Fig.~\ref{tree}(b), considering a variable node with degree $l$ (i.e., UE corresponding to this variable node sends $l$ sub-pilots to the BS),  an edge connecting to a variable node with degree $l$ can be recovered   when at least one of the other $l-1$ edges have been recovered. Let $n(m)$ denote the probability that an edge connecting to a factor (variable) node with degree $l$ cannot be recovered. Then, we have
\begin{equation}\label{23}
n = 1-\left(1-m\right)^{l-1},
\end{equation}
and
\begin{equation}\label{24}
m = n^{l-1}.
\end{equation}

By averaging
$m$ and $n$ over the edge distributions,
the evolution of the average erasure probability during the
$i^{th}~(1 \le i \le I)$ iteration where $I$ is the maximum number of iterations, can be written as
\begin{equation}\label{25}
\begin{array}{rcl}
n_i&=&\sum\nolimits_l {\rho_l \left(1 - \left(1 - m_{i-1}\right)^{l-1}\right)}\hfill \\  \\
& =& 1-\rho \left(1-m_{i-1}\right) ,\hfill \\  \\
m_i &=& \sum\nolimits_l {\lambda_l n_i^{l-1}} = \lambda \left(n_i\right).
\end{array}
\end{equation}
The update process of $m_i$ and $n_i$ is shown in Fig.~\ref{tree}(c). Specifically, given initial values $m_0$ and $n_0$, $m_i$ and $n_i$  can be derived by (\ref{25}).


It can be seen from Fig.~\ref{tree}(b) that,   a variable node with degree
$l$ is unknown when all of these $l$ edges cannot be recovered in
the $I^{th}$ iteration. So, the probability of the variable node being
unknown is $(n_I)^l$. By averaging $(n_I)^l$ over the degree distributions
of variable nodes, $P_\text{fail}$ can be calculated by~\cite{pfail}

\begin{equation}\label{26}
P_\text{fail}=\sum\nolimits_l {\Lambda_l(n_I)^l}=\Lambda (n_I).
\end{equation}
Substituting~(\ref{26})~into~(\ref{17}), we have
\begin{equation}\label{27}
S_\text{SIC}=N_a \left(1-P_\text{fail}\right)=N_a(1-\Lambda (n_I)).
\end{equation}
\subsubsection{$S_{\rm{CICA}}$ computation }\label{ICA number}

\

 The number of successful UEs obtained by  the CICA algorithm can be written as
\begin{equation}\label{28}
S_\text{CICA} =(N_a-S_\text{SIC})P_{\text{CICA}}^s,
\end{equation}
where $P_{\text{CICA}}^s$ denotes the probability that the CSI of one UE can be obtained successfully.

 Based on the procedure of the CICA algorithm  described in Subsection~\ref{CICA}, we can see that,  the CICA algorithm first obtains the decoded uplink message via (\ref{14}), and then obtains UE's CSI  via (\ref{15}). Thus, if the uplink message of one UE can be decoded successfully,  the CSI of this UE can be recovered successfully. Then, $P_{\text{CICA}}^s$ can be computed by
\begin{equation}\label{29}
P_{\text{CICA}}^s = \left(1-P_e\right)^{N_\text{m}}.
\end{equation}
where $P_e$ is the bit error rate (BER) of the proposed CICA algorithm, which is larger than that using maximum likelihood (ML) decoding method \cite{ML}, and can be written as

\begin{align}\nonumber
{P_e} &\ge \dfrac{1}{{\sqrt \pi  }}{\int}_{\sqrt {\rho_1 \beta_1 \sum\nolimits_{m = 1}^M {{{({g_{1,m}})}^2}} } }^{\infty}  {{e^{ - {x^2}}}dx} \hfill \\\nonumber \\ \label{30}
&\mathop  = \limits^{(a)} \dfrac{1}{{\sqrt \pi  }}{\int}_0^\infty  {
	{\int}_{\sqrt {y} }^\infty  {\dfrac{{{y^{M/2 - 1}}{e^{ - y/2}}}}{{{2^{M/2}}\Gamma (\frac{M}{2})}}}{e^{ - {x^2}}} } dxdy,
\end{align}
where step (a) follows from the fact that $\sum\nolimits_{m = 1}^M {{{({g_{1,m}})}^2}}$ is a central chi-squared distribution with degree of freedom $M$ and that ${\rho_1 \beta_1}=1$, and  $\Gamma (z)$ is the gamma function with parameter $z$ which is given by
\begin{equation}\label{31}
\Gamma (z) = \int_0^\infty  {{t^{z - 1}}{e^{ - t}}dt}.
\end{equation}

Substituting~(\ref{30})~into~(\ref{29}), we can derive the upper bound for $P_{\text{CICA}}^s$,
\begin{equation}\label{32}
P_{\text{CICA}}^s\le P_{\text{CICA}}^{\text{U}}= \resizebox{0.65\hsize}{!}{$\left(1-\dfrac{1}{{\sqrt \pi  }}{\int}_0^\infty  {
		{\int}_{\sqrt {y} }^\infty  {\dfrac{{{y^{M/2 - 1}}{e^{ - y/2}}}}{{{2^{M/2}}\Gamma (\frac{M}{2})}}}{e^{ - {x^2}}} } dxdy \right)^{N_\text{m}}$},
\end{equation}
and thus derive the upper bound for $S_\text{CICA}$
\begin{align}\nonumber
S_\text{CICA} &\le S_\text{CICA}^{\text{U}}  \\ \nonumber
&=P_{\text{CICA}}^{\text{U}} (N_a-S_\text{SIC}) \hfill \\ \nonumber
&=\resizebox{0.75\hsize}{!}{$\left(1-\dfrac{1}{{\sqrt \pi  }}{\int}_0^\infty  {
		{\int}_{\sqrt {y} }^\infty  {\dfrac{{{y^{M/2 - 1}}{e^{ - y/2}}}}{{{2^{M/2}}\Gamma (\frac{M}{2})}}}{e^{ - {x^2}}} } dxdy \right)^{N_\text{m}}$}  \\ \label{33}
&\quad \times (N_a-S_\text{SIC}).
\end{align}

Substituting~(\ref{33}), (\ref{27})~into~(\ref{16}), we can get the upper bound for the  successful access
probability of the proposed GCICA-RA scheme
\begin{align}\nonumber
P_{s}\le P_{s}^{\text{U}}&=\frac{ \resizebox{0.75\hsize}{!}{$(N_a\Lambda(n_I))\left(1-\dfrac{1}{{\sqrt \pi  }}{\int}_0^\infty  {
		{\int}_{\sqrt {y} }^\infty  {\dfrac{{{y^{M/2 - 1}}{e^{ - y/2}}}}{{{2^{M/2}}\Gamma (\frac{M}{2})}}}{e^{ - {x^2}}} } dxdy \right)^{N_\text{m}}$}}{N_a} \\ \label{34}
	&\qquad +\frac{N_a(1-\Lambda (n_I))}{N_a}.
\end{align}

\subsection{Missed detection probability analysis}

The missed detection probability refers to the ratio of the number of undetected active UEs to the total number of active UEs, which can be calculated as
\begin{equation}\label{35}
P_\text{md} = \dfrac{N_a -S_\text{GCICA-RA}}{N_a} =1-P_{s}.
\end{equation}

Substituting~(\ref{34})~into~(\ref{35}), we can get the lower bound for the missed detection probability
\begin{equation}\label{36}
\begin{array}{rcl}
P_{\text{md}} &\ge& P_{\text{md}}^{\text{L}} \hfill \\
&=&1-P_{s}^{\text{U}} \hfill \\
&=&1-[\frac{N_a(1-\Lambda (n_I)) }{N_a}+(N_a\Lambda(n_I))\times \\ &\qquad& \frac{\resizebox{0.75\hsize}{!}{$\left(1-\dfrac{1}{{\sqrt \pi  }}{\int}_0^\infty  {
		{\int}_{\sqrt {y} }^\infty  {\dfrac{{{y^{M/2 - 1}}{e^{ - y/2}}}}{{{2^{M/2}}\Gamma (\frac{M}{2})}}}{e^{ - {x^2}}} } dxdy \right)^{N_\text{m}}$}}{N_a}].
\end{array}
\end{equation}
\subsection{Uplink throughput analysis}
We define the uplink throughput as follows
 \begin{equation}\label{37}
 \begin{array}{l}
  \gamma  = \dfrac{{{S_\text{SchemeName}}{N_\text{PD}}R}}{{N_\text{PD}}+ O_\text{SchemeName}}
  =\dfrac{{{N_a}{P_s}{N_\text{PD}}R}}{{N_\text{PD}}+ O_\text{SchemeName}},
  \end{array}
\end{equation}
where ${S_\text{SchemeName}}$  denotes the  number of successful UEs where  subscript ``$\text{SchemeName}$" stands for the name of the scheme, $O_\text{SchemeName}$ denotes the overhead of the random access  scheme, and ${N_\text{PD}+ O_\text{SchemeName}}$ is the frame length.  Fig.~\ref{frame-structure} indicates  that  the amount of total overhead of the proposed GCICA-RA scheme is  $O_\text{GCICA-RA}=\tau_p \times L+1$.

Substituting~(\ref{34})~into~(\ref{37}), we can derive the upper bound for the uplink throughput of the proposed GCICA-RA scheme
\begin{equation}\label{38}
  \gamma\le \gamma^{\text{U}}=\dfrac{{{N_aP_{s}^{\text{U}}}{N_\text{PD}}R}}{{N_\text{PD}}  + \tau_p \times L+1}.
\end{equation}

\section{Numerical Results}

In this section, we first present the performance of  the proposed GCICA-RA scheme, including the successful access probability,  missed detection probability, and MSE performance of the CSI estimation method. Then, we make a comparison  between  the proposed GCICA-RA scheme, multiple-preamble grant-free
RA scheme proposed in~\cite{MultiplePreambles} and the  traditional random access scheme in terms of the uplink throughput. For the traditional random access scheme, active UEs sends their randomly selected pilots along with their uplink messages to the BS, and the BS can only decode the uplink message of UE free from pilot collision.

In the simulation, we consider the urban micro scenario, where the  path loss exponent is 3.8~\cite{Spatialchannel}.
 The radius of the cell is 200 meters and all UEs are uniformly distributed among the location farther than 25 meters from  the BS. Simulation parameters are shown in Table \ref{tab:parameters}. In addition, we utilize a LDPC code with 1/2 rate to code the payload data,  and then employ the BPSK scheme to module the coded payload data. In the simulation, since the length of payload data   is  a few hundred bytes in  M2M communications~\cite{byte}, the payload data size is set  to 128 bytes to meet the communication requirements of M2M communications, and further to  verify the proposed GCICA-RA scheme.

 \begin{table}[htbp]
\scriptsize
\centering
 \caption{Simulation parameters}\label{tab:parameters}
   \begin{tabular}{|c|c|}
\hline
\linespread{2}
Parameter & Value\\
\hline
$R$~ (Code rate) & 1/2\\
\hline
$N_\text{PD}$~ \tabincell{c}{(Symbol length of the coded \\ payload data)}& 2048\\
\hline
$M$~(The number of antennas)& 400 \\
\hline
 $\text{SNR}_e$~ \tabincell{c}{(Median signal to noise ratio (SNR) \\ of the UEs at the corner of cell)}& 5, 10, 15, 20 dB\\
\hline
$\tau_p$~(The number of pilots during each sub-pilot phase)& 6,9,10\\
\hline
$L$~(The number of sub-pilot phases)& 2,3\\
\hline
$N_{\rm{I}}$~(The number of ICA classifiers)& 5$\sim$30\\
\hline
$N_a$~\tabincell{c}{(The number of active UEs during \\ one random access procedure)} & 10 $\sim$ 40\\
\hline
\end{tabular}
\end{table}

\begin{figure}[htbp]
  \centering
\includegraphics[scale=0.5]{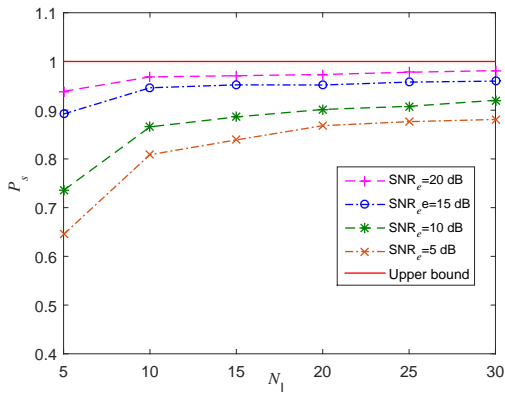}\\
 \caption{   Successful access probability versus the number of ICA classifiers $N_\text{I}$.}\label{success}
\end{figure}

\begin{figure}[htbp]
\centering
 \includegraphics[scale=0.5]{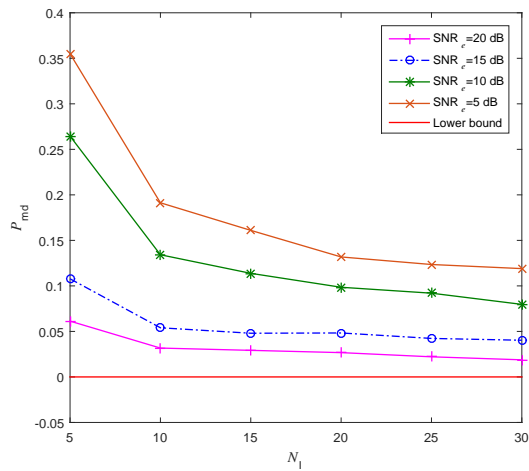}\\
  \caption{Missed detection probability versus the number of ICA classifiers $N_\text{I}$.}\label{unsuccess}
\end{figure}

\begin{figure}[htbp]
\centering
 \includegraphics[scale=0.5]{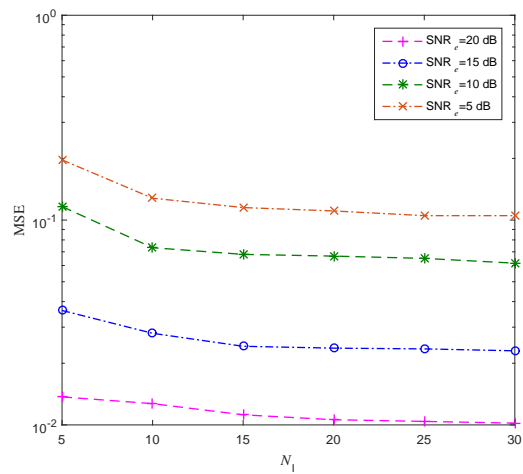}\\
  \caption{MSE performance versus the number of ICA classifiers $N_\text{I}$.}\label{MSE}
\end{figure}


\begin{figure}[htbp]
\centering
\includegraphics[scale =0.5] {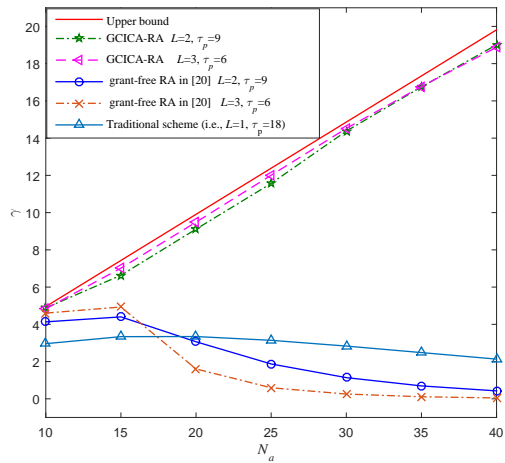}\\
\caption{Uplink throughput comparison between GCICA-RA and multiple-preamble grant-free RA scheme.}\label{throughput}
\end{figure}

Fig.~\ref{success} shows how the successful access probability  $P_s$ changes with the number of ICA classifiers $N_\text{I}$, where $\text{SNR}_{e}=5, 10, 15, 20$~dB, $M=400$, $L=2$, $N_{a}=20$ and $\tau_p=10$. The upper bound for the successful access probability can be obtained according to (\ref{34}).  We can see  from  Fig.~\ref{success} that,  with the increase of $N_\text{I}$, the successful access probability  increases dramatically  first and then increases at a slower pace for different $\text{SNR}_{e}$. The reason is that,  the more the number of ICA classifiers, the more signals of one UE we can obtain, which further improves the decoding performance. We can also observe that, with the increase of $\text{SNR}_{e}$, the successful access probability  increases, and  is close to the upper bound.

Fig.~\ref{unsuccess} shows how the missed detection probability changes with the number of ICA classifiers $N_\text{I}$, where $\text{SNR}_{e}=5, 10, 15, 20$~dB, $M=400$, $L=2$, $N_{a}=20$ and $\tau_p=10$. The lower bound for the missed detection probability can be obtained according to (\ref{36}).  Fig.~\ref{unsuccess} illustrates that,  with the increase of $N_\text{I}$, the missed detection probability  decreases dramatically  first and then decreases at a slower pace for different $\text{SNR}_{e}$. We can also note that, with the increase of $\text{SNR}_{e}$,  the missed detection probability takes small values and decreases, and is close to the lower bound.

Fig.~\ref{MSE} illustrates  MSE performance of the proposed GCICA-RA, where $\text{SNR}_{e}=5, 10, 15, 20$~dB, $M=400$, $L=2$, $N_{a}=20$ and $\tau_p=10$. We can see from Fig.~\ref{MSE} that, with the increase of $N_\text{I}$, MSE  decreases dramatically  first and then decreases at a slower pace for different $\text{SNR}_{e}$. We can also note that MSE takes small values and decreases with the increase of $\text{SNR}_{e}$.

Fig.~\ref{throughput} compares the uplink throughput between the proposed GCICA-RA,  multiple-preamble grant-free RA schemes in~\cite{MultiplePreambles}, and the  traditional random access scheme with different $L$. We set $M=400$, $\text{SNR}_{e}=10$ dB  and  $N_\text{I}=30$. To see  how the uplink throughput changes with the number of sub-pilot phases, we fix the length of the super pilot to 18, i.e., $L\times \tau_p=18$, and set $L$  to 2 and 3.
To compare  fairly, we ensure all random access scheme have the same pilot length and the same symbol length of the modulated uplink message.
 The upper bound for the uplink throughput of  the proposed GCICA-RA is obtained via (\ref{38}).  Note that, since upper bounds for the uplink throughput of the proposed  GCICA-RA are same  for cases of $L=2$ and  $L=3$  based on  (\ref{38}), we  not distinguish these two upper bounds in this figure.
 It is observed that, the uplink throughput of   $L=3$ is slightly higher than that of $L=2$ for the proposed  GCICA-RA scheme. The reason is that the number of iterations of the SIC for $L=2$  is larger than that for $L=3$ on average, which  results in the higher CSI estimation error~\cite{SUCRGBPA} and further increase  the equivalent noise introduced by the SIC algorithm.
  We also see that, the uplink throughput of the proposed GCICA-RA   scheme is close to the upper bound, and significantly higher than  those of  baselines. With the increase of the  number of active UEs, the  gap between the uplink throughput of the proposed GCICA-RA scheme and those of baselines increases.  This means that our proposed GCICA-RA is more suitable to  crowded scenarios.
\section{Conclusion}

An GCICA-RA  grant-free scheme was proposed to support massive connectivity for  M2M communications in massive MIMO systems.  The GCICA-RA scheme allows each active UE to transmit their super pilots and modulated uplink messages to the BS. Based on the received pilot signal and modulated uplink message, the BS  utilizes  a proposed GCICA algorithm to estimate the CSI and decode the uplink message of each active UE jointly, and then employs the estimated CSIs to  detect active UEs.  Simulation results  demonstrated that,  our proposed GCICA-RA  scheme  achieves performance gain even in the very crowded scenarios.


\end{document}